\documentclass[twocolumn,aps,preprintnumbers,prl,showpacs,amsfonts,amsmath,amssymb,superscriptaddress,floatfix]{revtex4}
\usepackage[dvips]{graphicx}
\usepackage[]{verbatim}
\allowdisplaybreaks[3]
\begin{document}
\title{Demonstration of Quadrature Squeezed Surface-Plasmons in a Gold Waveguide}
\date{\today}
\author{Alexander Huck}\email{alexander.huck@fysik.dtu.dk}
\affiliation{Department of Physics, Technical University of Denmark, Building 309, 2800 Lyngby, Denmark}
\author{Stephan Smolka}
\affiliation{DTU Fotonik, Department of Photonics Engineering, Technical University of Denmark, \O rsted Plads 343, 2800 Lyngby, Denmark}
\author{Peter Lodahl}
\affiliation{DTU Fotonik, Department of Photonics Engineering, Technical University of Denmark, \O rsted Plads 343, 2800 Lyngby, Denmark}
\author{Anders S. S\o rensen}
\affiliation{QUANTOP, Danish Quantum Optics Center and Niels Bohr Institute, 2100 Copenhagen, Denmark}
\author{Alexandra Boltasseva}
\affiliation{DTU Fotonik, Department of Photonics Engineering, Technical University of Denmark, \O rsted Plads 343, 2800 Lyngby, Denmark}
\affiliation{Birck Nanotechnology Center, Purdue University, West Lafayette 47907, IN, USA}
\author{Jiri Janousek}
\affiliation{Department of Physics, Technical University of Denmark, Building 309, 2800 Lyngby, Denmark}
\author{Ulrik L. Andersen}\email{ulrik.andersen@fysik.dtu.dk}
\affiliation{Department of Physics, Technical University of Denmark, Building 309, 2800 Lyngby, Denmark}

\begin{abstract}
We report on the efficient generation, propagation, and re-emission of squeezed long-range surface-plasmon polaritons (SPPs) in a gold waveguide. Squeezed light is used to excite the non-classical SPPs and the re-emitted quantum state is fully quantum characterized by complete tomographic reconstruction of the density matrix. We find that the plasmon-assisted transmission of non-classical light in metallic waveguides can be described by a Hamiltonian analogue to a beam splitter. This result is explained theoretically. 
\end{abstract}

\maketitle
Enormous interest has recently been devoted to the emergent field of quantum plasmonics due to its unique capabilities in the way electromagnetic radiation can be localized and manipulated at the nanoscale. In particular, integrated quantum technologies based on surface plasmons hold great promises for quantum information processing, since it allows for scalability, miniaturization, and coherent coupling to single emitters~\cite{2006Chang,2007Chang,2007Fedutik, 2007Akimov,2003Bergman}. To enable these quantum information processing technologies with high fidelity, it is of paramount importance, that the nonclassicality of the plasmonic modes is preserved in propagation. The first experiment verifying the preservation of entanglement in plasmonic nanostructures was carried out by Alterwischer et al.~\cite{Plasmon_Entangle}. They demonstrated the survival of polarization entanglement after plasmonic propagation through subwavelength holes in a metal film. The preservation of energy time entanglement in a perforated metal film as well as in a thin conducting waveguide was later demonstrated by Fasel et al.~\cite{2005Fasel}. These experiments have witnessed the preservation of probabilistically prepared entanglement (thus neglecting the vacuum contribution) described in the two dimensional Hilbert space.

In the present Letter we investigate the compatibility of the quantum plasmonic technology with the continuous variable quantum domain (described in the infinite dimensional Hilbert space) by demonstrating the plasmonic excitation, propagation, and detection of deterministically prepared quadrature squeezed vacuum states. We show that a squeezed vacuum state excite an electron resonance on the surfaces of a metallic gold waveguide to form a surface plasmon polariton (SPP). Despite loss and decoherence in the plasmonic mode we demonstrate that quadrature squeezing is retained in the retrieved light state. Importantly, we fully characterize the input state and output state by performing a complete quantum tomographic reconstruction of the states density matrix. This is in strong contrast to previous experiments on plasmon assisted quantum state transmission, where only a certain property of the quantum state was investigated. 

SPPs are combined electron oscillations and electromagnetic waves propagating along the interface between a conductor and a dielectric medium~\cite{1986Burke}. By reducing the thickness of the metal film the SPP from the upper interface and the lower interface can couple. This coupling results in the formation of either long-range (LR) SPPs or short-range (SR) SPPs depending on the propagation length of the mode~\cite{1986Burke}. In general, the coupled LR-SPP (SR-SPP) modes are characterized by an increased (decreased) propagation length compared to a SPP propagating along a single interface. LR-SPPs (SR-SPPs) are further characterized by a symmetric (anti-symmetric) longitudinal electric field distribution, which allows the LR-SPPs to be efficiently excited using end-fire coupling~\cite{2000Berini, 2005Boltasseva}.

\begin{figure}[htbp]
\includegraphics[width=0.48\textwidth]{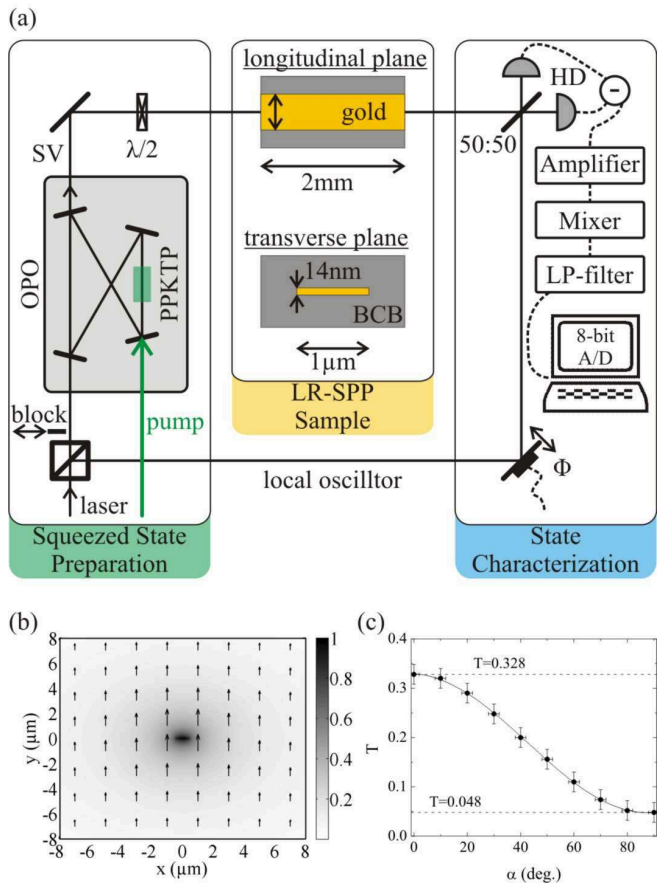}
\caption{(Color online). (a) Sketch of the experimental setup and LR-SPP sample: The sample is made of gold embedded in lossless transparent BCB and its dimensions are specified as written in the figure. Experimental setup: OPO - optical parametric oscillator, PPKTP - periodically poled KTP crystal, SV - squeezed vacuum, $\lambda/2$ - half-wave plate, $\Phi$ - piezo actuated mirror for phase variation, 50:50 - symmetric beam splitter, and HD - homodyne detection scheme. (b) Normalized power density plot of the LR-SPP mode for sample dimensions as specified in (a). The arrows show the direction of the electric field in the transverse plane. At the experimental wavelength of $1064nm$ the refractive indices of gold and BCB are specified to be $n_{Au}=0.2381+i7.7199$~\cite{1972Johnsen} and $\text{n}_{\text{BCB}}\text{=1.539}$, which were used in the simulation. (c) Dependence of the total transmission T on the angle $\alpha$ of the linear polarized incident light field. $\alpha$ is measured with respect to TM polarization and set by the $\lambda/2$ wave plate. \label{sample}}
\end{figure}

A sketch of the sample as well as the experimental setup is shown in Fig.~\ref{sample}(a). The sample consists of a gold stripe embedded in lossless transparent polymer Benzocyclobutene (BCB). It was prepared by first spinning a 12-15$\mu m$ layer of BCB onto a silicon substrate. The metallic waveguide was then patterned using standard UV lithography, followed by gold deposition by means of electron beam evaporation and lift-off. Finally, another layer of BCB with a thickness of 10 $\mu m$ was spun onto the sample with exactly the same spinning conditions to ensure a symmetric dielectric environment of the gold stripe. In Fig.~\ref{sample}(b) we present a power density plot of the LR-SPP mode in the transverse plane of the gold stripe. This was obtained by a finite element method simulation for the metal stripe geometry as presented in Fig.~\ref{sample}(a). The arrows in Fig.~\ref{sample}(b) show the direction of the electric field vector in the transverse plane of the sample, which is, with respect to the metal stripe surface, transverse magnetic (TM) polarized. This fact is in agreement with the analytical model for a sample of infinite width~\cite{1986Burke}. 

As a source of non-classical light we use a bow-tie shaped optical parametric oscillator (OPO) operating below threshold, see Fig.~\ref{sample}(a)~\cite{2006Suzuki}. The non-linear medium inside the OPO is a type-III periodically poled KTP (PPKTP) crystal of 10mm length, which is placed at the beam waist between two curved highly reflective mirrors with 25mm radius of curvature. We pump the PPKTP crystal with a frequency doubled Nd:YAG laser field having a wavelength of 532nm, thus producing squeezed light centered around 1064nm. The round trip path length of the OPO cavity is 275mm, it has a free spectral range of 1.1GHz, and a bandwidth of approximately 21MHz. The out-coupling mirror has a power transmissivity of T=10\%. 

Before investigating the quantum properties of a SPP, we characterize its classical properties. For this purpose, we seed the OPO cavity with an auxiliary light field at 1064nm with the OPO pump beam blocked. A half wave plate is placed between the OPO and the gold sample to control the linear polarization of the light field. The light field is then focused onto the sample with an aspheric lens ensuring a good mode matching to the LR-SPP mode. At maximum, we have observed a total transmission of $\text{32.8}\pm0.5\%$ for TM polarized incident light. When rotating the polarization of the incident light field, the transmission decreases as $\cos^{2}(\alpha)$, where $\alpha$ is the angle of the half-wave plate with respect to the TM direction. For $\alpha=90^{\circ}$ only scattered as well as higher order modes are measured. This result clearly demonstrates the expected polarization dependence and thus proves, that the LR-SPP mode is excited~\cite{2000Berini}. 

\begin{figure}[htbp]
\includegraphics[width=0.48\textwidth]{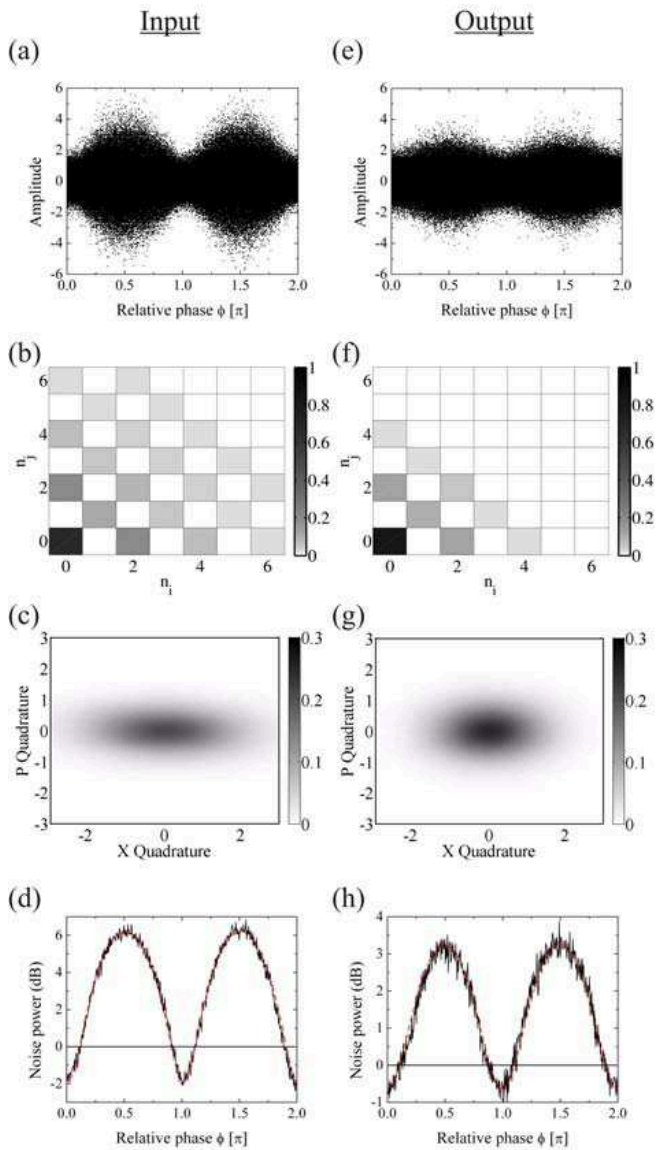}
\caption{(Color online). Experimental results of the squeezed vacuum state (input) and the quantum state after LR-SPP excitation, propagation, and reemission (output). (a) and (e) Time resolved data for various electric field quadratures while scanning the relative phase between LO and signal from 0 to $\text{2}\pi$. The visibility between signal light field and LO is $88.5\%$ in case of the squeezed vacuum field and $90.3\%$ in case of the LR-SPP mode. The data are normalized in a basis, where $\left[\hat{x},\hat{p}\right]=i$. (b) and (f) Reconstructed density matrices $\rho_{in}$ and $\rho_{LR-SPP}$ in the photon number basis obtained by applying the maximum likelihood algorithm to the data presented in (a) and (e), respectively. $\rho$ has been reconstructed for a truncated Hilbert space with a photon number of n=15. (c) and (g) Calculated Wigner functions $W(X,P)$ for the density matrices shown in (b) and (f), respectively. (d) and (h) Noise power of the data set presented in (a) and (f) (black solid line) and calculated from the density matrices (red dashed line).
\label{results}}
\end{figure}

Having established the classical excitation of a SPP, we now proceed with the analysis of non-classical SPPs. In the following we fully characterize the quantum state of light before it is launched into the gold stripe, and also after transmission and re-emission. The results of the analysis are shown in Fig.~\ref{results}. For the characterization of the input state, the sample is removed from the setup and the squeezed vacuum light field is measured with a homodyne detection system, as shown in Fig.~\ref{sample}(a). The difference signal of the two detectors is amplified, mixed down at an optical sideband frequency of 4.7MHz, low pass filtered with a bandwidth of 150kHz, and finally digitized with 8bit resolution. A quadrature measurement set containing more than 5.5M data points is shown in Fig.~\ref{results}(a) for a rotation in phase space from 0 to $\text{2}\pi$. By applying the maximum likelihood algorithm~\cite{MaxLik} on this data set we reconstruct the squeezed vacuum states density matrix $\rho_{in}$, whose absolute values are presented in Fig.~\ref{results}(b). From the density matrix $\rho_{in}$, the Wigner function $W(X,P)$ has been reconstructed, which is presented in Fig.~\ref{results}(c). For various qudratures, the noise power can either be calculated directly from the time resolved data, as shown by the black solid line in Fig.~\ref{results}(d), or from the states density matrix $\rho_{in}$, as shown by the red dashed line in Fig.~\ref{results}(d). The degree of squeezing and anti-squeezing are found to be $\text{-1.9} \pm \text{0.1dB}$ and $\text{6.1}\pm\text{0.1dB}$, respectively.

The squeezed light is then carefully injected into the gold stripe, thereby linearly mapping the non-classical photon statistics onto a plasmonic state, as described for the exciation with a single photon in Ref.~\cite{2008Tame}. Subsequently, the squeezed SPP propagates through the sample, is efficiently re-emitted at the rear end, and measured with the homodyne detector. The time resolved quadrature data are recorded similarly as was described above and the result is shown in Fig.~\ref{results}(e). From these data we reconstruct the density matrix $\rho_{LR-SPP}$ and the Wigner function $W(X,P)$, as illustrated in Fig.~\ref{results}(f) and (g), respectively. By calculating the quadrature variances (either directly from the time-resolved data or from the density matrix) we find a minimum of $\text{-0.7}\pm\text{0.1dB}$ and a maximum of $\text{3.2}\pm0.1\text{dB}$ with respect to the shot noise level, as is presented in Fig.~\ref{results}(h). We therefore conclude, that the retrieved light state is squeezed and that the squeezing survived the plasmonic propagation.    

\begin{figure}
\includegraphics[width=0.44\textwidth]{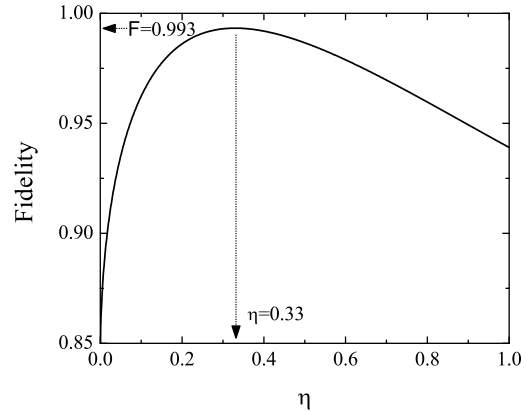} \vspace*{-0.5cm}
\caption{Fidelity between beam splitter mode $\rho_{out}(\eta)$ obtained after applying the beam splitter operator $\hat{U}_{BS} (\eta)$ on the input mode $\rho_{in}$ and LR-SPP output mode $\rho_{LR-SPP}$ versus transmissions $\eta$. \label{fidelity}}
\end{figure}

We furthermore investigate whether the operation, that transforms the density matrix of the input state $\rho_{in}$ to that of the output state $\rho_{LR-SPP}$ can be described by the unitary beam splitter operator $\hat{U}_{BS}=exp\{ \frac{\theta}{2}(\hat{a}^{\dagger}\hat{b}e^{i\phi} - \hat{a} \hat{b}^{\dagger} e^{-i\Phi})\}$. Here, $\hat{a}$ and $\hat{b}$ are the field operators of the beam splitter input modes, $\Phi$ is the relative phase between the modes $\hat{a}$ and $\hat{b}$, and $\theta$ is linked to the transmission $\eta$ via $\eta=cos(\frac{\theta}{2})$. The expected output state is thus $\rho_{out}(\eta)=Tr\{U_{BS}(\eta)\rho_{in}\otimes|0\rangle\langle 0|U_{BS}^\dagger(\eta)\}$, where the trace is taken over one of the output modes of the beam splitter. To see whether $\rho_{out}(\eta)$ is similar to the actually measured output state $\rho_{LR-SPP}$ we compute the fidelity $F(\eta)$ between the two states. The fidelity is given by $F(\eta)=tr\{\sqrt{\rho_{out}(\eta)} \rho_{LR-SPP} \sqrt{\rho_{out}(\eta)}\}^{1/2}$~\cite{ref:fid}, with $0\leq F(\eta)\leq 1$ and $F(\eta)=1$ if and only if $\rho_{out}(\eta)=\rho_{LR-SPP}$. In Fig.~\ref{fidelity}, we plot the fidelity $F(\eta)$ versus the transmission $\eta$ of the beam splitter, which reaches $F=0.993$ at maximum for $\eta=0.33$. The large overlap between $\rho_{out}(\eta=0.33)$ and $\rho_{LR-SPP}$ are in excellent agreement with the measurements presented in Fig.~\ref{sample}(c) and proves, that the plasmonic decoherence is linear and thus can be simulated by a beam splitter interaction.

In the following we justify this conclusion by using simple theoretical arguments. From the quantum mechanical Maxwell equations the propagation of the field in an (electrically) polarizable medium may in general be described by 
\begin{equation}
{\left (\frac{\partial^2}{\partial t^2}+\frac{1}{c^2} \nabla\times\nabla\times\right)}\hat E=-\frac{1}{\epsilon_0} \frac{\partial^2}{\partial t^2} \hat P,
\label{eq:propagation}
\end{equation}
where $\hat E$ is the electric field and $\hat P$ is the polarization of the medium. We are here interested in the component at optical frequencies and split the electric field into the positive and negative frequency components $\hat E=\hat E^{(+)}+\hat E^{(-)} ,$ where the positive (negative) frequency component contains the photon  annihilation (creation) operator. Similarly, we split the polarization $\hat P=\hat P^{(+)}+\hat P^{(-)}$. The positive frequency component of the polarization may be expanded in terms of the electric field
\begin{equation}
\hat P^{(+)} = \sum_{m=0}^\infty\sum_{n=0}^\infty \hat c_{mn}(\{\hat{O}_i\}) {\left({\hat E^{(-)}}\right)}^m {\left({\hat E^{(+)}}\right)}^n  ,
\label{eq:expansion}
\end{equation}
where we include the possibility that the expansion coefficient $\hat c_{mn}$ may depend on a set of state operators $\{\hat{O}_i\}$ describing the state of the polarizable medium. During the propagation, the photonic excitations of the electric field are mapped onto the (quasiparticle) excitations of the polarizable medium. The expansion coefficient $c_{mn}(\{\hat{O}_i\})$ thus gives access to information about the nature of the excitation of the medium and, e.g., quasi particle interactions. Our experiment is performed in the low intensity limit, in which case the polarization reduces to $P^{(+)}=\hat c_{00}(\{\hat{O}_i\})+\hat c_{01}(\{\hat{O}_i\}) \hat E^{(+)}$, where we have excluded the $\hat E^{(-)}$ term since a negligible intensity is observed when the incoming field is in vacuum. This also means that we must require $\langle\hat c_{00}^\dagger \hat c_{00}\rangle=0$. Since the equation of motion (\ref{eq:propagation}) is linear, the expression for $\hat P^{(+)}$ means that the outgoing field $\hat E_{{\rm out}}^{(+)}$ can be written as a combination of two terms $\hat E_{{\rm out}}^{(+)}=\hat G(\{\hat{O}_i\}) \hat c_{00}(\{\hat{O}_i\})+\hat G(\{\hat{O}_i\}) \hat E_{{\rm in}}^{(+)}$, where $\hat G(\{\hat{O}_i\})$ is the Greens function of the plasmonic propagation. If we  ignore any dependence on internal state operators $\{\hat{O}_i\}$ in the expansion coefficient $\hat c_{00}$ and $\hat c_{01}$ in Eq. (\ref{eq:expansion}), we find that the input/output relation for a single mode operator $\hat a$ is  given by 
\begin{equation}
\hat a_{{\rm out}}= \sqrt{1-\eta} \hat v + \sqrt{\eta }\hat a_{{\rm in}}, \label{eq:BS}
\end{equation}
where $\hat v$ is a combination of the $\hat c_{00}$ operators with the factor $\sqrt{1-\eta}$ separated out for convenience. From $\langle\hat c_{00}^\dagger \hat c_{00}\rangle=0$ then immediately follows that  $\langle \hat v^\dagger \hat v\rangle =0$ and consistency of the commutation relations requires that $[\hat v,\hat v^\dagger]=1$. $\hat v$ is thus a single mode vacuum operator in agreement with our experimental observation. Our results therefore show that the classical description $ P^{(+)} = (\epsilon - \epsilon_0) E^{(+)}$ (where the permittivity of the material $\epsilon$ is just a constant) remains valid for the metal down to the level of single photons if we add a vacuum contribution to the operator equations. This conclusion is consistent with the theoretical interpretation in Ref. \cite{2004moreno} of the experiments in Ref. \cite{Plasmon_Entangle}, where it is concluded that the polarization degrees of freedom leaves no "which-way" information in the solid. The present work, however, extends that conclusion by showing that also the presence or absence of a photon leaves no "which-way" information. 

In conclusion, we have experimentally demonstrated the efficient plasmon-assisted propagation of squeezed vacuum states in a gold waveguide. Through complete reconstruction of the density matrices of the input and output fields we found that the plasmonic propagation can be described by a unitary beam splitting operation. The squeezing was therefore coherently transferred from the light state to the plasmonic state and back to the light state, solely degraded by vacuum noise. This demonstrated robustness of continuous variable quantum states in plasmon propagation suggests that continuous variable quantum information processing based on surface plasmons is feasible. 

We gratefully acknowledge support from the Villum Kann Rasmussen Foundation, the Danish Research Council, and the EU project COMPAS.


\begin{thebibliography}{99}
\bibitem{2006Chang} D.E. Chang, A.S. S\o rensen, P.R. Hemmer, and M.D. Lukin, Phys. Rev. Lett. \textbf{97}, 053002 (2006), D.E. Chang, A.S. S\o rensen, P.R. Hemmer, and M.D. Lukin, Phys. Rev. B \textbf{76}, 035420 (2007).

\bibitem{2007Fedutik} Y. Fedutik, V.V. Temnov, O. Sch\"ops, U. Woggon, M.V. Artemyev, Phys. Rev. Lett. \textbf{99}, 136802 (2007).

\bibitem{2007Akimov} A.V. Akimov et al., Nature (London) \textbf{450}, 402 (2007).

\bibitem{2007Chang} D.E. Chang, A.S. S\o rensen, E.A. Demler, and M.D. Lukin, Nature Physics (London) \textbf{3}, 807 (2007).

\bibitem{2003Bergman} D.J. Bergman and M.I. Stockman, Phys. Rev. Lett. \textbf{90}, 027402 (2003).

\bibitem{Plasmon_Entangle} E. Altewischer, M.P. van Exter, J.P. Woerdman, Nature (London) \textbf{418}, 304 (2002).

\bibitem{2005Fasel} S. Fasel, F. Robin, E. Moreno, D. Erni, N. Gisin, and H. Zbinden, Phys. Rev. Lett \textbf{94},110501 (2005).

\bibitem{1986Burke} J.J. Burke, G.I. Stegeman, and T. Tamir, Phys. Rev. B \textbf{33}, 5186 (1986).

\bibitem{2000Berini} R. Charbonneau, P. Berini, E. Berolo, and E. Lisicka-Shrzek, Optics Letters \textbf{25}, 11 (2000).

\bibitem{2005Boltasseva} A. Boltasseva et al., Journal of Lightwave Technology, \textbf{23}, 1 (2005).

\bibitem{1972Johnsen} P.B. Johnsen and R.W. Christy, PRB \textbf{6}, 4370 (1972).

\bibitem{2006Suzuki} S. Suzuki, H. Yonezawa, F. Kannari, M. Sasaki, and A. Furusawa, Appl. Phys. Lett. 89, 061116 (2006).

\bibitem{2008Tame} M.S. Tame, C. Lee, J. Lee, D. Ballester, M. Paternostro, A.V. Zayats, and M.S. Kim,Phys. Rev. Lett. \textbf{101}, 190504 (2008).

\bibitem{MaxLik} A.I. Lvovsky, Journal of Optics B \textbf{6}, 556 (2004).

\bibitem{ref:fid} R. Jozsa, Journal of Modern Optics \textbf{41}, 2315 (1994).

\bibitem{2004moreno} E. Moreno, F.J. Garc\'ia-Vidal, D. Erni, J.I. Cirac, and L. Mart\'in-Moreno, Phys. Rev. Lett. \textbf{92}, 236801 (2004).

\end{thebibliography}
\end{document}